\documentclass[conference]{IEEEtran}
\IEEEoverridecommandlockouts
\usepackage{subcaption}
\usepackage{csquotes}
\usepackage[nolist,nohyperlinks]{acronym}
\usepackage{cite}
\usepackage{amsmath,amssymb,amsfonts}
\usepackage{algorithmic}
\usepackage{graphicx}
\usepackage{textcomp}
\usepackage{xcolor}
\usepackage{balance}
\def\BibTeX{{\rm B\kern-.05em{\sc i\kern-.025em b}\kern-.08em
    T\kern-.1667em\lower.7ex\hbox{E}\kern-.125emX}}

\usepackage{cleveref}

\begin{document}

%\title{Likable or intelligent? Comparing social robots and virtual agents for long-term health monitoring}

\title{Likable or Intelligent? Comparing Social Robots and Virtual Agents for Long-term Health Monitoring}

% \title{Likable or Intelligent? Comparing Social Robots and Virtual Agents for Long-term Health Monitoring
% \thanks{This research was partially supported by the German Federal Ministry of Education and Research (BMBF) in the projects GeneRobot (project number 16GDI102A) and HINT (project number 16SV9182).}
% }

% \title{Likable or intelligent? Comparing social robots and virtual agents for long-term health monitoring
% \thanks{This research was partially supported by the German Federal Ministry of Education and Research (BMBF) in the projects GeneRobot (project number 16GDI102A) and HINT (project number 16SV9182).}
% }

\begin{acronym}[TOPICS-SF]
\acro{AAL}{ambient assisted living}
\acro{AB}{AdaBoost}
\acro{ADL}{activities of daily living}
\acro{BADL}{basic activities of daily living}
\acro{IADL}{instrumental activities of daily living}
\acro{AI}{artificial intelligence}
\acro{ANN}{artificial neural network}
\acro{API}{Application Programming Interface}
\acro{app}{application}
\acro{ASQ}{After-Scenario-Questionnaire}
\acro{ASR}{automatic speech recognition}
\acro{ATI}{affinity for technology interaction}
\acro{BERT}{bidirectional encoder representations from transformers}
\acro{BMI}{body mass index}
\acro{BPS}{biopsychosocial}
\acro{BPSE}{biopsychosocialenvironmental}
\acro{CB}{CatBoost}
\acro{CCL}{Cologne Cobots Lab}
\acro{CGA}{Comprehensive Geriatric Assessment}
\acro{COPD}{chronic obstructive pulmonary disease}
\acro{DHA}{Digital Health Assessments}
\acro{DL}{deep learning}
\acro{DSR}{design science research}
\acro{DSRM}{design science research methodology}
\acro{DT}{decision trees}
\acro{EDA}{exploratory data analysis}
\acro{ECA}{embodied conversation agent}
\acro{EKG}{electrocardiogram}
\acro{ePROMs}{electronic PROMs}
\acro{ERD}{Entity-Relationship-Diagram}
\acro{FEDS}{Framework for Evaluation in Design Science}
\acro{FN}{false negative}
\acro{FP}{false positive}
\acro{GB}{gradient boosting}
\acro{GCPS}{Graded Chronic Pain Scale}
\acro{GDPR}{General Data Protection Regulation}
\acro{GOHAI}{Geriatric Oral Health Assessment Index}
\acro{GPAQ}{Global Physical Activity Questionnaire}
\acro{GP}{general practitioner}
\acro{GPT}{generative pre-trained}
\acro{GQS}{Godspeed questionnaire series}
\acro{HLS}{health literacy survey}
\acro{HRQoL}{health-related quality of life}
\acro{HRV}{heart rate variability}
\acro{HRI}{Human-Robot Interaction}
\acro{H-TAM}{Healthcare Technology Acceptance Model}
\acro{iADL}{instrumental activities of daily living}
\acro{ICOPE}{Integrated Care for Older People}
\acro{IoT}{Internet of Things}
\acro{IS}{information systems}
\acro{IT}{information technology}
\acro{IVA}{intelligent virtual agent}
\acro{KNN}{k-nearest neighbors}
\acro{KPI}{Key Performance Indicator}
\acro{LGBM}{LightGBM}
\acro{LLM}{large language model}
\acro{LR}{logistic regression}
\acro{MCC}{majority class classifier}
\acro{MCI}{mild cognitive impairment}
\acro{MET}{metabolic equivalent}
\acro{ML}{machine learning}
\acro{MLP}{multilayer perceptron}
\acro{MNA-SF}{Mini Nutritional Assessment Short Form}
\acro{MPI}{Multi-dimensional Prognostic Index}
\acro{NB}{Naïve Bayes}
\acro{NLP}{natural language processing}
\acro{PCA}{principal component analysis}
\acro{PHQ-4}{Patient Health Questionnaire-4}
\acro{PROM}{patient-reported outcome measure}
\acro{PROMs}{Patient-Reported Outcome Measures}
\acro{RAG}{retrieval-augmented generation}
\acro{RF}{random forest}
\acro{REST}{representational state transfer}
\acro{RQ}{research question}
\acro{RSES}{Rosenberg Self-Efficacy Scale}
\acro{SAR}{socially assistive robot}
\acro{SDK}{software development kit}
\acro{SELFY-MPI}{Self-administered Multi-dimensional Prognostic Index}
\acro{SAS}{Smiley-Analog-Scale}
\acro{SDG}{sustainable development goals}
\acro{SIA}{socially interactive agent}
\acro{SRC}{stratified random classifier}
\acro{SC}{stacking classifier}
\acro{STT}{Speech-To-Text}
\acro{SUS}{System Usability Scale}
\acro{SVM}{support vector machine}
\acro{TAM}{Technology Acceptance Model}
\acro{TN}{true negative}
\acro{TOPICS-SF}{The Older Persons and Informal Caregivers Survey Short Form}
\acro{TP}{true positive}
\acro{UEQ}{User Experience Questionnaire}
\acro{UEQ+}{modular extension of the User Experience Questionnaire}
\acro{UI}{User Interface}
\acro{UN}{United Nations}
\acro{UTAUT}{Unified Theory of Acceptance and Use of Technology}
\acro{UX}{user experience}
\acro{VA}{virtual agent}
\acro{VC}{voting classifier}
\acro{WHO}{World Health Organization}
\acro{WOz}{Wizard of Oz}
\acro{XAI}{eXplainable artificial intelligence}
\acro{XGB}{XGBoost}
\end{acronym}

%%% For camera-ready version
\author{\IEEEauthorblockN{Caterina Neef}
\IEEEauthorblockA{\textit{Cologne Cobots Lab} \\
\textit{TH Köln - University of Applied Sciences}\\
Cologne, Germany\\
caterina.neef@th-koeln.de}
\and
\IEEEauthorblockN{Anja Richert}
\IEEEauthorblockA{\textit{Cologne Cobots Lab} \\
\textit{TH Köln - University of Applied Sciences}\\
Cologne, Germany\\
anja.richert@th-koeln.de}}
%%%

%%% For peer review version
% \author{
% \IEEEauthorblockN{Anonymous authors}
% \IEEEauthorblockA{\textit{Name of department} \\
% \textit{Name of organization}\\
% City, Country \\
% email address or ORCID}}
%%%

\maketitle

\begin{abstract}
% The use of social robots and virtual agents as interfaces for health monitoring systems for older adults offers the possibility of more engaging interactions with systems that can support long-term health and well-being. 
% %They can help assess and monitor various health domains, such as quality of life, frailty, mental health, or physical activity. 
% %Each of these interfaces has its advantages:
% While robots are characterized by their physical presence, software-based virtual agents are more scalable and flexible.
% Few comparisons of these interfaces exist in the human-robot and human-agent interaction domains, especially in long-term and real-world studies. 
% In this work, we examined impressions of social robots and virtual agents at both the beginning and end of an eight-week study in which older adults interacted with these systems independently in their own homes.
% Using a between-subjects design, participants could choose which of these interfaces they preferred to use as the front end for health monitoring during the study. 
% While the social robot was perceived as somewhat more likable, the \ac{VA}was perceived as more intelligent. 
% Our work highlights the need for further studies investigating factors which are most relevant for engaging interactions with social interfaces for long-term health monitoring. 
Using social robots and virtual agents (VAs) as interfaces for health monitoring systems for older adults offers the possibility of more engaging interactions that can support long-term health and well-being. 
While robots are characterized by their physical presence, software-based VAs are more scalable and flexible.
Few comparisons of these interfaces exist in the human-robot and human-agent interaction domains, especially in long-term and real-world studies. 
In this work, we examined impressions of social robots and VAs at the beginning and end of an eight-week study in which older adults interacted with these systems independently in their homes.
Using a between-subjects design, participants could choose which interface to evaluate during the study. 
While participants perceived the social robot as somewhat more likable, the VA was perceived as more intelligent. 
Our work provides a basis for further studies investigating factors most relevant for engaging interactions with social interfaces for long-term health monitoring. 
\end{abstract}


\section{Introduction}
% The problem statement and research gap
Goal 3 of the United Nations' \ac{SDG} is to \enquote{ensure healthy lives and promote well-being for all at all ages}~\cite[para.~1]{UNSDGS}. %  add page or paragraph
To achieve this, addressing the health and care needs of older adults and promoting healthy aging is vital. 
One option to facilitate this are \acp{SIA}, such as social robots and \acp{VA}, which can support older adults in monitoring their own health~\cite{boumans2020,lanhingting2021,murali2023,jiang2024,neef2024a}.
Social robots are a promising approach in this context, as they lead to more positive interactions when delivering healthcare instructions, including asking health-related questions, compared to a computer tablet~\cite{mann2015}.
While a social robot's physical embodiment can lead to more user engagement, \acp{VA} are more mobile and available anytime and on the go, enabling very frequent and intimate interactions~\cite{bickmore2022}. 
As social robots and \acp{VA} each have their advantages and disadvantages, comparisons of both types of \acp{SIA} may shed light on which agent is best suited for which context and which user group.
However, such comparisons in the areas of human-robot and human-agent interaction remain scarce, especially in long-term and real-world studies~\cite{jung2018,park2020,trinh2018}.

\begin{figure*}[htbp]
     \centering
     \begin{subfigure}[b]{0.49\textwidth}
         \centering
         \includegraphics[width=\textwidth]{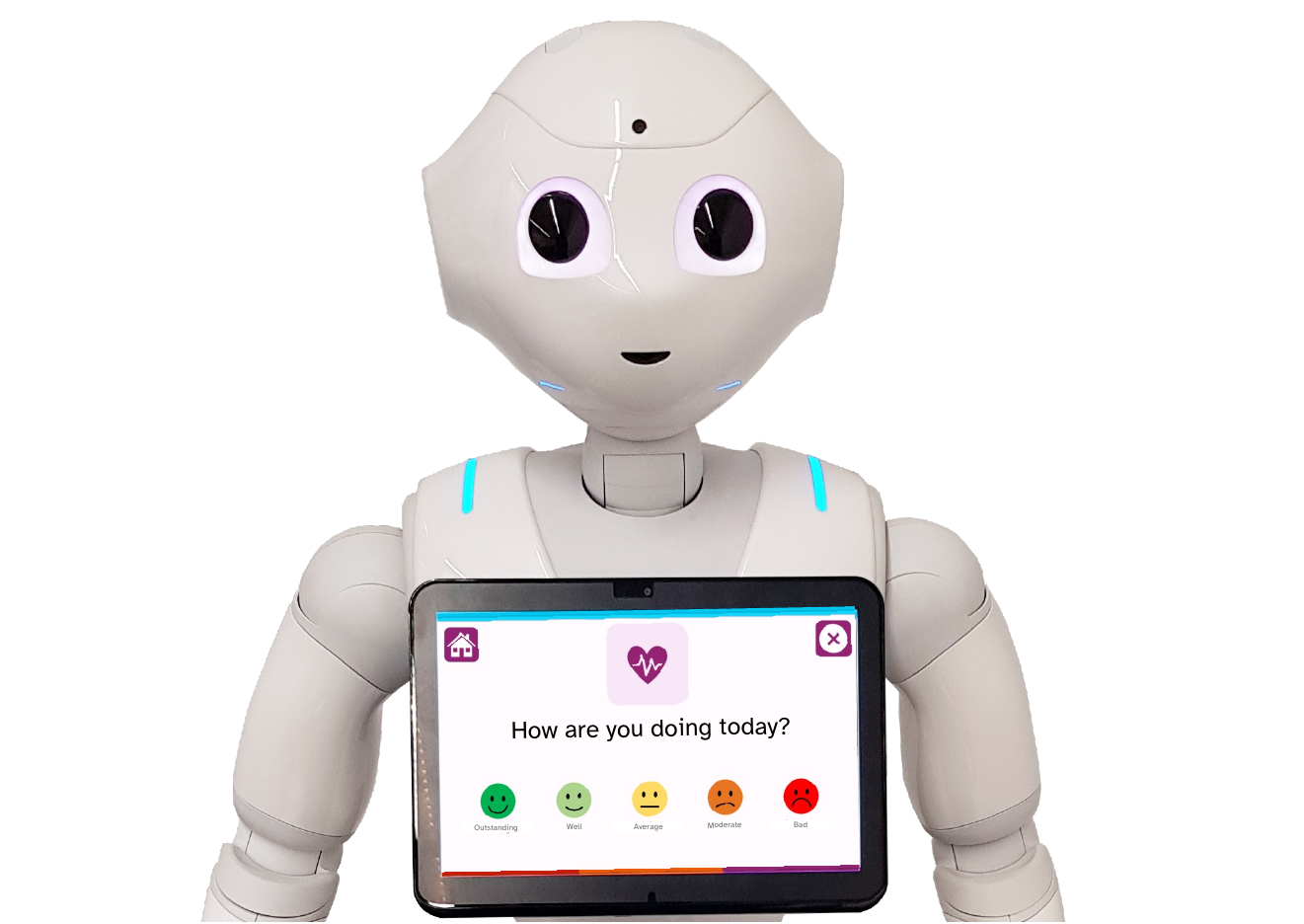}
         
         \caption{Pepper asking one of the daily questions.}
         \label{fig:pepper-daily}
     \end{subfigure}
     \hfill     
     \begin{subfigure}[b]{0.49\textwidth}
         \centering
         \includegraphics[width=\textwidth]{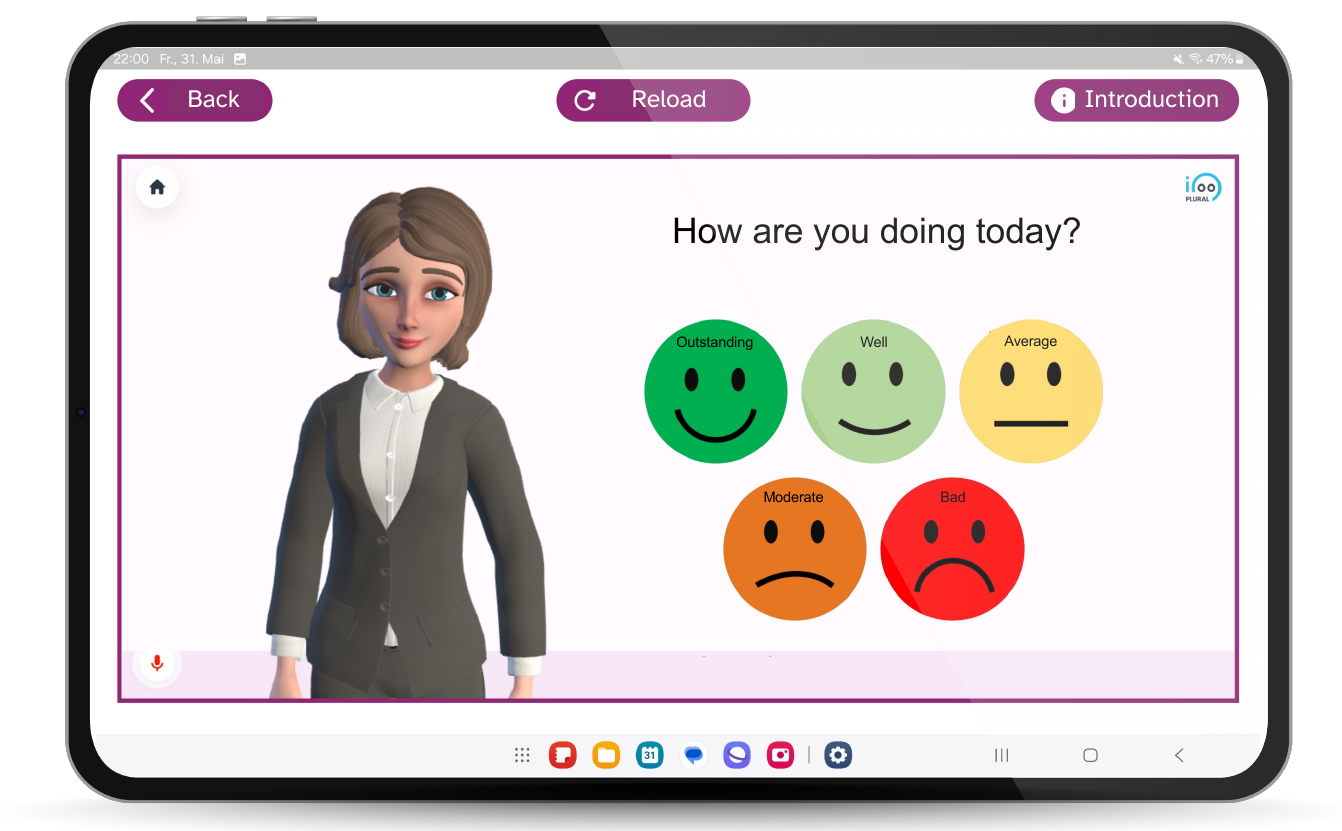}
         \caption{Charlotte asking one of the daily questions.}
         \label{fig:charlotte-daily}
     \end{subfigure}
        \caption{An example is shown of the social robot Pepper and the \ac{VA} Charlotte asking one of the health-related questions in this study, i.e., \enquote{How are you doing today?}. Screenshots were translated into English.}
        \label{fig:example-dailies}
\end{figure*}

% objective and contribution of the work
To this end, we present a comparison of the anthropomorphism, animacy, likability, and perceived intelligence using the \ac{GQS}, of a social robot and a \ac{VA} at the beginning and end of the study, in which we deployed the \acp{SIA} for eight weeks in older adults' homes. 
Using a between-subjects design, participants independently chose their preferred \ac{SIA} to evaluate. 
Both deployed agents, the social robot Pepper~\cite{pandey2018} (Aldebaran, France) and the \ac{VA} Charlotte~\cite{plural2024} (Humanizing Technologies, Germany/Austria), are shown in Fig.~\ref{fig:example-dailies}.
Participating older adults were asked to interact daily with the respective \ac{SIA}, which asked health- and well-being-related questions and conducted medical questionnaires with the older adults.
The \acp{SIA} also informed them of the results of the questionnaires, and were able to provide further information and tips regarding various health domains, e.g., quality of life, frailty, mental health, or physical activity. 
This work contributes one of the first systematic, longitudinal comparisons of older adults' evolving perceptions of a social robot and a \ac{VA} in a real-world health monitoring context. 
%With these findings, we highlight the need for further investigations of such comparisons to fully leverage each system's advantages. 
In the following, we present literature related to the use and comparison of social robots and \acp{VA} for health-related applications, describe the study and its results, and discuss implications gained from this research. 

\section{Related Work}
\label{sec:rel-work}
Robots are frequently used for health monitoring tasks, compared to screen-based interactions, as their physical presence as interaction partners can lead to more favorable responses and a more positive perception~\cite{li2015}.
In~\cite{gava2020}, a robot and a \ac{VA} were compared for small talk conversations among adults aged 45-65. 
The robot was considered friendlier and more trustworthy, while the \ac{VA} was favored for its learnability and interaction speed. 
Social robots are also a suitable choice for the health assessment of older adults in particular, as they can help combat social isolation in hospital settings~\cite{sarabia2018}.
According to a scoping review~\cite{gasteiger2021}, social robots can also help combat loneliness, but more research needs to be conducted on \acp{VA} in this domain. 
In an evaluation of both a \ac{VA} and a social robot to support isolated older adults~\cite{sidner2018}, participants preferred the robot in the home environment. 
However, the authors argue that \acp{VA} may help support older adults on the go, e.g., when getting exercise or going shopping, and that social robots and \acp{VA} may thus complement each other. 

In a comparison of the virtual and physical representation of the iCat robot with a text interface in the context of health self-management of older adults~\cite{looije2010}, the virtual and physical character were seen as more empathetic and trustworthy than the text-based interface, and the older adults were more conversational with the characters. 
Research on user preference for either a \ac{VA} or a robot has explored several influencing factors: 
In~\cite{nishio2021}, the advantages of the robot's physical presence for older users are highlighted, suggesting increased engagement in a coaching context for mental support. 
However, the used \ac{VA} was a replica of the robot, limiting facial expressions and thus compromising its human-like appeal. 
In~\cite{esposito2022}, older adults preferred interacting with \acp{VA} over robots, citing the \ac{VA}'s less robotic appearance and higher scores in reliability, practicality, and engagement.

While acceptance and willingness to use social robots and \acp{VA} for daily assistance are important factors, long-term motivation to actually use the agent is not guaranteed. 
When deploying \acp{SIA} in long-term studies in real-world environments, the novelty effect~\cite{smedegaard2019} needs to be considered, as the motivation and excitement can decline as soon as the first week of usage~\cite{ostrowski2022}. 
True long-term interactions may also be referred to as interactions that have gone beyond the point in time when the novelty effect wears off~\cite{leite2013}. 
Here, differences between individual users are also crucial to consider, as preferences, expectations, and comfort with technology can vary widely among older adults~\cite{durick2013, righi2017, schlomann2022}. 

% Research gap 
The current state of research suggests that user preferences for either \acp{VA} or robots are influenced by various subtle factors, often dependent on the specific application domain. 
However, previous studies have primarily focused on either highly realistic \acp{VA}~\cite{esposito2022} or virtual representations of the robot~\cite{nishio2021}. 
These approaches may not fully exploit the potential of \acp{VA}: 
Overly realistic \acp{VA} may create high user expectations that are difficult to meet~\cite{benmimoun2012}, while designing the \ac{VA} as a replica of the robot means inheriting the robot's limitations, such as the inability to display facial expressions.
As such, while each interface has its advantages and disadvantages, direct comparisons of both are still lacking in literature, particularly in real-world settings and contexts, and over longer periods of time, when the novelty of the interaction with the \ac{SIA} has worn off. 
To address this gap, we present an eight-week, in-home comparison of a social robot and a \ac{VA}, in which older adults chose their preferred interface for health monitoring.
\section{Materials and Methods}
For participation in this study, we recruited 23 older adults (defined by the United Nations as persons over the age of 60~\cite{unitednationshighcommissionerforrefugeesunhcr2020}) interested in self-managing their health using assistive technologies, through a local newspaper. 
11 participants, 6 of whom identified as female, with an average age of 71.3 years, chose to evaluate the robot (group R). 
They lived in 7 households, as some participants were couples who lived together and shared a robot. 
12 participants, 8 of whom identified as female, with an average age of 70.8 years, preferred to evaluate the \ac{VA} (group VA). 
They lived in 9 households, as this group also included couples living in the same household. 
The study was conducted in German; therefore all participants were required to be proficient German speakers.
Participants were informed in detail about the study procedures and purpose, and the collected and pseudonymized data in compliance with local data privacy regulations. %, as well as where the data was stored, and who had access to it.
% Todo blind before submission 
% Data privacy and informed consent were obtained from all participants, and this study was reviewed and approved by the Institution Review Board of the TH Köln - University of Applied Sciences Cologne (application nos. THK-2022-0004 and THK-2024-0002). 
Data privacy and informed consent were obtained from all participants, and this study was reviewed and approved by the Institutional Review Board of the TH Köln -- University of Applied Sciences. 
No compensation was provided to participants. 

To mimick realistic consumer choice for future real-world deployments of such systems, participants could decide which \ac{SIA} they wanted to evaluate. 
They were asked to interact daily with their respective \ac{SIA}, which asked questions about their health and well-being, as shown in Fig.~\ref{fig:example-dailies}. 
It also conducted medical questionnaires regarding the domains of health-related quality of life, frailty, physical activity, pain, mental health, and self-efficacy. 
One participant quit the study after 5 weeks due to personal reasons unrelated to the study. 
Participants interacted on between 22 and 56 days with the \acp{SIA}. 

The system builds on the health monitoring architecture introduced in~\cite{neef2022,neef2024a,neef2024b}.
We chose the social robot Pepper (Fig.~\ref{fig:pepper-daily}), as it has been successfully used in many studies in the healthcare domain~\cite{boumans2020,carros2022,fattal2022}, and a \ac{VA} (Fig.~\ref{fig:charlotte-daily}), as they have also been successfully used in the healthcare domain~\cite{bickmore2018,jiang2024,murali2023}.
We used an Android-based Pepper with its included QiChat dialog manager, and a \ac{VA} running on the plural platform~\cite{plural2024}, with the conversational \ac{AI} framework Rasa~\cite{rasa2022} (USA) as a dialog manager.
The platform plural offers cartoon-like \ac{VA} characters of different genders and races, suitable to avoid high expectations that may follow highly realistic appearances of \acp{VA}~\cite{benmimoun2012}.
As research has shown, older adults prefer female \acp{VA}~\cite{esposito2019}, and there are indications that they may prefer agents which are not very young, as this may convey inexperience~\cite{terstal2020,terstal2020a}. 
Therefore, we chose the character \textit{Charlotte} (shown in Fig.~\ref{fig:charlotte-daily}) due to her gender, perceived age and outfit, which we hoped would evoke a sense of competence in users. 
The dialogs implemented in both systems were the same and were conducted autonomously by the agents, and we designed the \acp{UI} of both applications with accessibility factors in mind, especially for the target group of older adults who may have impaired vision or hearing~\cite{erharter2016}. 

To evaluate the perception of the social agent, we used the \ac{GQS}~\cite{bartneck2023}, which is one of the most highly cited and used questionnaires in the field of \ac{HRI}. 
It measures the perception of an agent's anthropomorphism, animacy, likability, its perceived intelligence, and perceived safety. 
Each of these measures can be used individually, and includes between three and six items, based on semantic differentials and a five-point scale with opposing items. 
We asked participants to fill out the \ac{GQS} items, randomized in their order, using a paper-based questionnaire at the beginning and end of the eight-week study. %, when we set up the system and when we returned to retrieve it. 
We opted not to analyze the perceived safety items, as participants feedbacked that it was unclear which point in time these items referred to.
Our objective was to compare the two \acp{SIA}, as well as a possible change in perception of the \acp{SIA} after eight weeks. 
We chose the eight week study period to reduce the effect of the system novelty on the results~\cite{smedegaard2019,leite2013,ostrowski2022}. 
We explained how to use the system to participants, and were available throughout the study duration for technical support. 
This investigation was part of an overall larger study, incorporating additional questionnaires regarding e.g., \ac{UX}, as well as bi-weekly interviews for additional qualitative feedback. 
We present the \ac{GQS} results from this study in the following. %, including a t-test to investigate the significance.% of the results. 
\section{Results}
\begin{figure*}[h]
     \centering
     \begin{subfigure}[b]{0.495\textwidth}
         \centering
         \includegraphics[width=\textwidth]{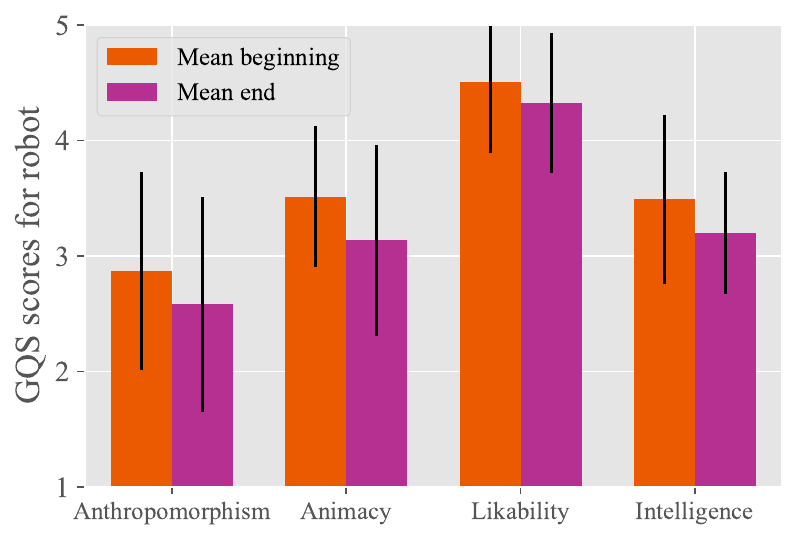}
         \caption{The \ac{GQS} results for the robot.}
         \label{fig:pepper-gqs-results}
     \end{subfigure}
     \hfill
     \begin{subfigure}[b]{0.495\textwidth}
         \centering
         \includegraphics[width=\textwidth]{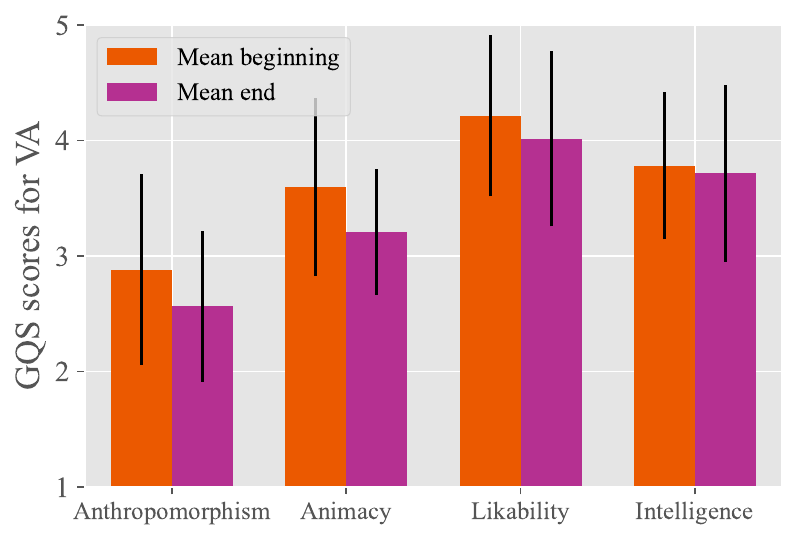}
         \caption{The \ac{GQS} results for the \ac{VA}.}
         \label{fig:charlotte-gqs-results}
     \end{subfigure}     
        \caption{The \ac{GQS} scores for the agents Pepper (a) and Charlotte (b) are visualized at the beginning and end of the eight-week study, with minimum values corresponding to 1 and maximum values to 5. All ratings are lower at the end of study, and anthropomorphism and animacy are rated similarly for both agents, while Pepper is perceived as slightly more likable and Charlotte is perceived as more intelligent.}
        \label{fig:gqs-results}
\end{figure*}

The scores of the anthropomorphism, animacy, likability, and perceived intelligence series of the \ac{GQS} questionnaire are visualized in Fig.~\ref{fig:gqs-results} for both group R rating Pepper and group \ac{VA} rating Charlotte, both at the beginning and end of the eight-week study. 
The minimum value corresponds to 1, the maximum to 5. 
Overall, Pepper and Charlotte were rated very similarly, and all ratings were lower by the end of the study, compared to the ratings at its beginning. 
Participants rated the anthropomorphism of both agents in the medium range and rated it slightly lower at the end of the study than at the beginning. 
The animacy was rated somewhat higher for both agents, and also decreased by the end of the study. 
Both systems were rated as very likable.
Pepper was rated as slightly more likable than Charlotte at 4.51 and 4.22 at the beginning of the study, respectively, although both ratings slightly decreased by the end of the study to 4.33 and 4.02. 

Regarding perceived intelligence, Pepper had a lower score than Charlotte, both at the beginning (3.49 vs. 3.78) and at the end of the study (3.20 vs 3.72).
We conducted a t-test; neither the before/after comparison of the composite scores of anthropomorphism, animacy, likability, and intelligence nor the comparison between Pepper and Charlotte were significant. %, according to a t-test we conducted. 
When looking at the individual items of the measures, however, we found that the items \textit{irresponsible/responsible} ($p = 0.006$) and \textit{unintelligent/intelligent} ($p = 0.045$), both part of the perceived intelligence measure, differed significantly in the comparison of the two \acp{SIA} at the end of the study, with responsibility values of 3.09 vs. 4 and intelligence values of 2.64 and 3.33 for Pepper and Charlotte, respectively. 

\section{Discussion and Conclusion}
The results of the \ac{GQS} questionnaires showed that while the anthropomorphism and animacy of Pepper and Charlotte were rated similarly, differences in likability and perceived intelligence emerged. 
Pepper was seen as slightly more likable, possibly due to its physical embodiment and thus more tangible presence, which can evoke the feeling of increased social presence~\cite{li2015,mann2015,gava2020}. 
In contrast, Charlotte was perceived as more intelligent, aligning with literature which suggests that specific design choices, such as a more mature appearance of the agent and professional attire, may evoke a feeling of competence and credibility in users~\cite{terstal2020,terstal2020a}. 

While our study is exploratory in nature and has only a limited sample size, these differences raise key questions regarding the role of embodiment and appearance of the social agents in shaping older adults' long-term engagement and acceptance of social health monitoring technologies. 
Factors to investigate here include whether likability or perceived intelligence lead to more sustained engagement with the system, or whether the defining factor is that users are able to choose the \ac{SIA} in its respective embodiment according to their preference, i.e., favoring likability and the corresponding \ac{SIA} over perceived intelligence, or vice versa.  
This underscores the importance of providing older adults with the option and the opportunity to choose their preferred \ac{SIA}.
Additionally, both virtually and physically embodied agents should be compared with, e.g., app-based health monitoring solutions, possibly with or without an included voice assistant for more interaction, to investigate the role embodiment plays.

Notably, all \ac{GQS} measures declined by the end of the study, suggesting that the initial enthusiasm of interacting with the system may wear off once users have become accustomed to it. 
This aligns with the previously described novelty effect~\cite{smedegaard2019,leite2013,ostrowski2022}, as the participants may have grown used to the \ac{SIA} as it became a routine part of their daily life. 
To address this, it is essential to understand how to mitigate this effect, e.g., through more personalization, adaptivity, or incremental feature introduction to keep the system interaction exciting. 
Potential strategies here include a larger variety of health-related questions and assessments tailored to the individual user's health status, along with personalized health feedback provided by the system. 

Once increased user engagement is achieved, future work should investigate whether this leads to positive effects on health outcomes in the long run, due to improved health awareness enabled by the health monitoring system. 
This is one of the main objectives of deploying such systems, to truly support healthy aging, and contribute to reaching the global goal of ensuring healthier lives and promoting well-being for all. 
Therefore, future research should examine whether an \ac{SIA}'s perceived intelligence, its likability, or its embodiment, lead to more consistent responses to the health questionnaires, which may lead to improved self-monitoring and better health management. 
Other factors, such as trust or humor, may also play a role in continuous use and should be examined.

Another research question to address in future work is what constitutes an optimal health monitoring companion for older adults, and how this may differ from one individual to another, as older adults represent a heterogeneous user group of technology~\cite{durick2013, righi2017, schlomann2022}. 
It would be interesting to examine in greater depth which factors influence perception and design of the agent in the context of health monitoring, and how these factors differ between individuals. 
Some may prefer a familiar, empathetic companion, while others may value efficiency, accuracy, or health expertise. 
Factors that may influence the perception of an agent's intelligence could include its appearance, personality, mobility, flexibility, facial expressions, gestures, and additional functions it should possess outside of the health monitoring domain. 
These may include small talk, the weather, or current news, if any, and wishes may vary widely between users.
Understanding these individual differences could guide the development of customizable \ac{SIA} profiles with dynamically adjustable or adaptive styles, appearances, and features, depending on user preference. 
This customization may also contribute to increased user engagement in the long-term. 

Ultimately, this exploratory study highlights the complexity of designing \acp{SIA} that effectively support older adults in monitoring their health. 
By examining the interplay of embodiment, appearance, likability, intelligence, and user perceptions over time, we take a step toward identifying factors that are most relevant for sustained engagement with health monitoring systems to facilitate technology-assisted healthy aging. 
Future research should build on these findings to develop adaptive and personalized systems that can truly contribute to healthier lives for all. 

\section*{Acknowledgments}
This research was partially supported by the German Federal Ministry of Education and Research (BMBF) in the projects GeneRobot (project no. 16GDI102A) and HINT (16SV9182).
The authors would like to thank the participants and Katharina Linden, Katinka Rosenfeld, Leon Munz, and Michael Macher for their invaluable contributions to this work. 
% %[BLINDED]

%\section*{References}
\balance
\bibliographystyle{IEEEtran} 
\bibliography{IEEEabrv,HRI2025-LBR-CNAR}

\end{document}